# Electronic transport and thermoelectric properties of phosphorene nanodisk under an electric field

M.Amir Bazrafshan, Farhad Khoeini*

*Department of Physics, University of Zanjan, P.O. Box 45195-313, Zanjan, Iran*

*Corresponding author

**Abstract**

The Seebeck coefficient is an important quantity in determining the thermoelectric efficiency of a material. Phosphorene is a two-dimensional material with a puckered structure, which makes its properties anisotropic. In this work, a phosphorene nanodisk (PDisk) with a radius of 3.1 nm connected to two zigzag phosphorene nanoribbons is studied, numerically, by the tight-binding (TB) and non-equilibrium Green's function (NEGF) methods in the presence of transverse and perpendicular electric fields. Our results show that the change of the structure from a zigzag ribbon form to a disk one creates an energy gap in the structure, so that for a typical nanodisk with a radius of 3.1 nm, the size of the energy gap is 3.88 eV. Besides, with this change, the maximum Seebeck coefficient increases from 1.54 to 2.03 mV/K. Furthermore, we can control the electron transmission and Seebeck coefficients with the help of the electric fields. The numerical results show that with the increase of the electric field, the transmission coefficient decreases, and the Seebeck coefficient changes. The effect of a perpendicular electric field on the Seebeck coefficient is weaker than a transverse electric field. For an applied transverse electric field of 0.3 V/nm, the maximum Seebeck coefficient enhances to 2.09 mV/K.

## 1. Introduction

Thermoelectric is an active field in nanotechnology to convert thermal energy into electricity. Researchers are trying to enhance the thermoelectric efficiency of various materials. In recent decades, owing to high precision tools, engineering the shape of the materials at nanoscales where quantum mechanical effects are playing a major role in determining the physical properties [1], is not just a dream.

After the realizing graphene in 2004 [2], the world of two-dimensional (2D) materials attracted the attention of researchers, since they possess extraordinary properties.

Phosphorene, the two-dimensional form of black phosphorus [3,4], is one of the materials that is of interest in the scientific community. Owing to its specific geometry, the physical properties of the phosphorene monolayer are not isotropic and depend on the direction [3–8]. Phosphorene is proposed for a variety of applications such as transistors, batteries, water splitting, sensors, and optoelectronic applications [9–12].

The chemical stability of black phosphorus is better than that of red and white ones [1]. Phosphorene is a direct bandgap semiconductor (>1.5 eV) with high carrier mobility [8,13–16]. Lattice thermal conductivity in the zigzag direction is higher than armchair direction for phosphorene [15,17]. Although, phosphorene is unstable when exposed to air. To overcome this challenge, usually it is sandwiched between other materials [16]. Tailoring nanostructures can change their physical properties, which for phosphorene, despite it is an intrinsic semiconductor, but its zigzag nanoribbon has metallic behavior [18].

Quantum confinement effects become important at the nanoscale, hence it can help control the physical properties of a material, including the Seebeck coefficient, which is essential for thermoelectric performance [19–21]. In this study, we investigate the electronic transport properties of a phosphorene nanodisk with a radius of R=3.1 nm connected to two zigzag phosphorene nanoribbons (ZPNR) with 12 atoms in the width (12-ZPNR), see Figure 1. We use the tight-binding and the non-equilibrium

Green's function methods for this goal. As reported in [22,23], the five hopping TB model can accurately reconstruct near the Fermi bands for phosphorene monolayer. We also apply two electric fields perpendicular to the transport direction up to 0.3 V/nm, close to the possible experimental values [24], to see how the system responds to transverse ($E_x$) and perpendicular ($E_z$) electric fields.

The manuscript arrangement is as follows; in the next section, we describe the model and method with a brief introduction to the TB and NEGF methods. In section three, the results and discussion are presented. In the last section, we conclude our study.

## 2. Model and Method

The system consists of a phosphorene disk connected to two metallic leads, as presented in Figure 1. Two zigzag phosphorene ribbon leads are labeled as source and drain. Besides, the central region between the two leads is a nanodisk device, which contains 798 atoms.

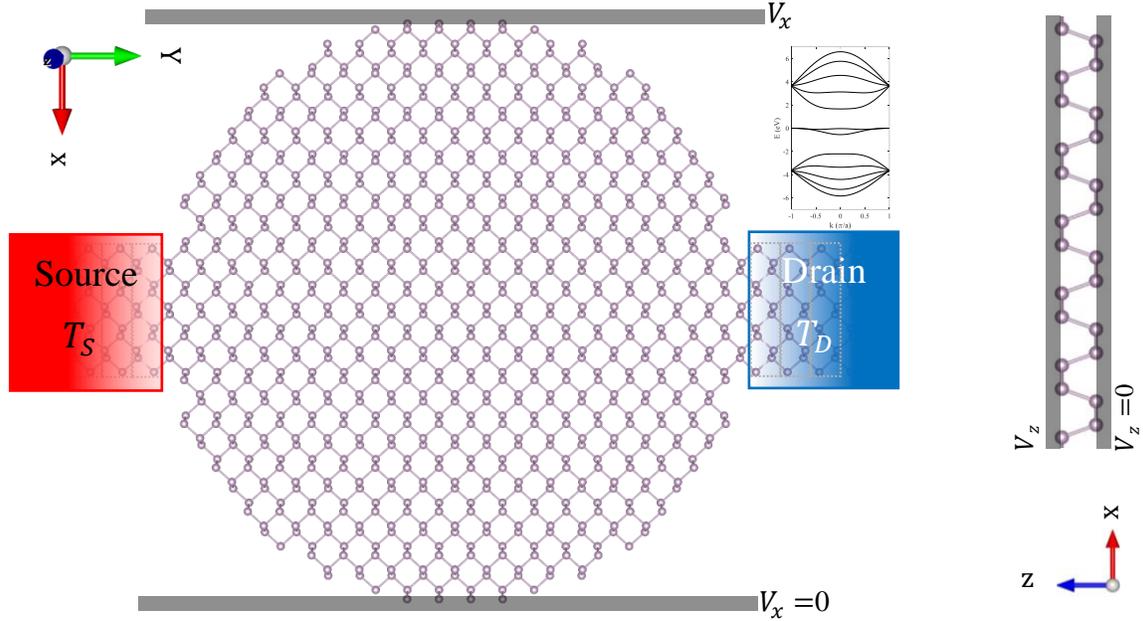

Figure 1. A phosphorene nanodisk with a radius of 3.1 nm connected to the two 12 ZPNR leads. The transport direction is along the y-axis. The source and drain leads act as the hot bath and the cold one, respectively. The TB electronic band structure of the 12-ZPNR is also shown on top of the drain lead. Transverse ($E_x$) and perpendicular ($E_z$) fields are only applied to the device placed between two leads (central region). Based on the relative position of the atoms to these lines, each atom feels a particular electric potential, which adds to its on-site energy. Electrode unit cells are shown by dotted rectangles.

The width of the ZPNR leads includes 12 atoms. Phosphorene is a semiconductor, but cutting it into a zigzag nanoribbon, turns it into a metal as mentioned in [18,25,26], and can be seen from the TB electronic band structure shown in Figure 1.

The electronic and transport behaviors of the system are described by the TB Hamiltonian:

$$H = \sum_{i} \varepsilon_i |i\rangle\langle i| + \sum_{<i,j>} (t_{i,j}|i\rangle\langle j| + \text{h.c.}),  \quad (1)$$

where $\varepsilon_i$ is the on-site energy and $t_{i,j}$ is the hopping parameter between atoms $i$ and $j$. The TB parameters are adopted from [23,27], which are $t_1 = -1.22, t_2 = 3.665, t_3 = -0.205, t_4 = -0.105$, and $t_5 = -0.055$ eV. Calculation method details can be found in [28]. The on-site energy, $\varepsilon_i$, is zero in the absence of an electric field, but it changes when the electric field is turned on. Here, according to Figure 1, in the presence of a transverse electric field, $E_x$, the on-site energy varies from zero to $\varepsilon_i = eE_x x_i$ ($x_i$ is the distance of the atom $i$ from the y-axis, x coordinate of atom $i$) [29]. This holds for $E_z$ as $\varepsilon_i = eE_z z_i$ with $z_i = 0$ or $z_i = d$, which $d$ is the thickness of phosphorene.

Moreover, the electric fields are only applied to the nanodisk, i.e., the device section. Therefore, the electrodes do not feel any electric field. The calculation method is available in [29]. To study the transport properties, the TB Hamiltonians are implemented in the NEGF formalism. The retarded Green's function can be evaluated as [30]:

$$G(E) = [(E + i\eta)I - H_C - \Sigma_{SC}(E) - \Sigma_{DC}(E)]^{-1}, \quad (2)$$

where $E$ is the electron energy, $I$ is the identity matrix, $\eta$ is an arbitrarily small positive number, $H_C$ is the central region (or device) Hamiltonian, and $\Sigma_{SC(DC)}$ is the self-energy for the source (drain) lead. Details about this formalism can be found in [16,31].

The spectral density operator is given by:

$$\Gamma_{S(D)}(E) = i[\Sigma_{SC(DC)}(E) - \Sigma_{SC(DC)}(E)^\dagger], \quad (3)$$

The transmission probability for the electron can be obtained as:

$$T(E) = \text{Trace}[\Gamma_S(E)G(E)\Gamma_D(E)G(E)^\dagger]. \quad (4)$$

Besides, the local density of state (LDOS) for a given atom (indicated by index $i$) can be derived by evaluating:

$$\text{LDOS}(E)_i = -\frac{1}{\pi}\text{Im}(G(E)_i), \quad (5)$$

Now, one can evaluate the electronic conductance, g, the Seebeck coefficient, S, and the electronic thermal conductance, κ, as [32,33]:

$$S(\mu, T) = \frac{1}{eT}\frac{L_1(\mu, T)}{L_0(\mu, T)}, \quad (6)$$

The elementary charge is indicated by $e$ in equation (6), and $L_n$ is given by:

$$L_n(\mu, T) = -\frac{2}{h}\int_{-\infty}^{\infty} T_e(E)\frac{(E-\mu)^n}{k_B T}\frac{\exp(\frac{E-\mu}{k_B T})}{\left(\exp(\frac{E-\mu}{k_B T}) + 1\right)^2} dE, \quad (7)$$

with $h$ as the Plank constant and $k_B$ as the Boltzmann constant. The temperature is $T = 300$ K in the calculations.

## 3. Results and discussion

The electronic transport properties and the Seebeck coefficient are studied in a phosphorene nanodisk with a radius of R=3.1 nm with the help of the TB method and NEGF formalism. As mentioned earlier, phosphorene is not an isotropic structure.

The electronic conductance in the zigzag direction is about an order of magnitude lower than the armchair direction, as reported in [6]. In contrast to electronic conductance, phononic conductance in the zigzag direction is about 40% higher than that of the armchair direction. As one can see in equations (6) and (7), the Seebeck coefficient is calculated based on the transmission coefficient. Hence to have a larger S, zero transmission coefficient with respect to the Fermi energy is of high importance (the wider energy gap, the higher S), or a transmission coefficient with sharp peaks followed by low values. Therefore, we first investigate the transmission coefficient. In the absence of any electric field, the ZPNR is a metal, as evidenced by its electronic band structure (Figure 1), and the transmission spectrum (thick red line in the panels of Figure 2). For the case of the nanodisk, the structure has an energy gap of 3.88 eV, as shown in Figure 2 A (magenta line). With the increase of the transverse electric field strength (blue line for $E_x$=0.15 V/nm, and green line for 0.3 V/nm), the electronic transmission coefficient of the system decreases. The effect of $E_z$ is shown in Figure 2 B . This field has a weaker effect on the transmission spectra, but still in the form of reduced transmission coefficient.

We note that colors are preserved in all figures respect to the structures and the studied electric fields. The behavior of the transmission coefficient is important in determining Seebeck coefficient. On the other hand, based on the formalism used in this work, at a fixed temperature, the overall of the transport coefficient is important in determining the Seebeck coefficient, as the integral in equation (7) says.

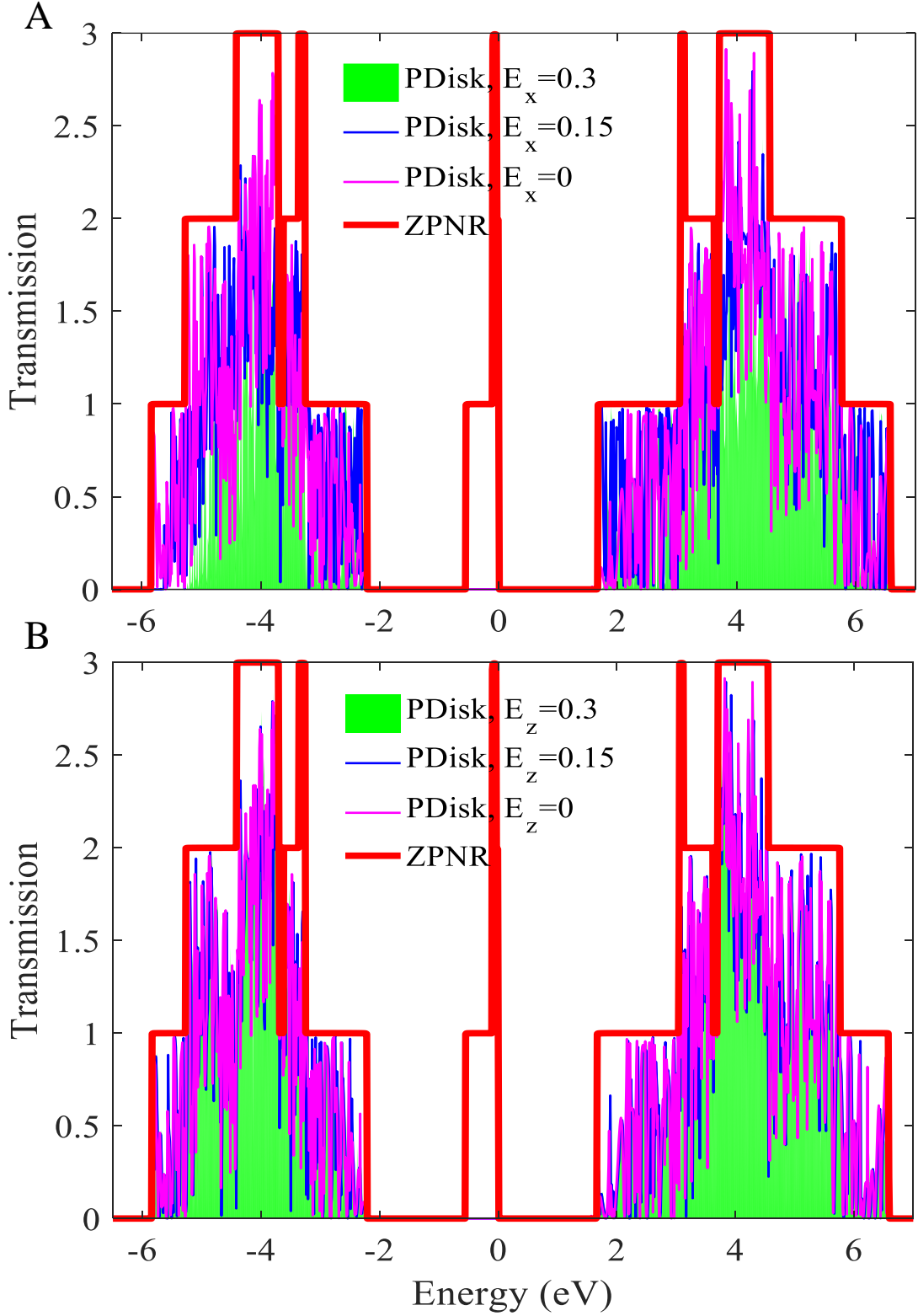

Figure 2. Transmission spectra of the phosphorene nanodisk, (A) in the presence of various $E_x$, (B) under the influence of $E_z$. The red line indicates transmission spectrum for the unbiased 12-ZPNR leads. Unit of the electric field is V/nm.

It is important to know what states are present in the device, so that the energy levels are plotted against the state index by solving the eigenvalue problem for each state in the device [34], see Figure 3 A . The bulk energy gap in the device, based on these energy levels, is ~1.66 eV, close to the phosphorene bandgap, indicating that from the electronic energy gap point of view, the behavior of this nanodisk is almost close to that of a single-layer phosphorene.

In the next step, we take a look into the local DOS. The normalized LDOS is obtained respect to the highest LDOS in the particular system. Figure 3 B presents the normalized LDOS of the PDisk with no bias. The LDOS shows that atoms in the upper and lower layers of the puckered structure of phosphorene have different local electron densities, and according to the central point of the device, there is a mirror symmetric behavior from the LDOS point of view in all directions. Here, we arrange the structure according to the origin of the coordination system (0, 0, 0). The high LDOS states are located where the electrodes are connected to the device. In the disk region, the LDOS distribution is not too different, a clue of lower localization, and thus, better conductivity. Higher localized states can enhance the Seebeck coefficient by reducing the contribution of electrons in the conductivity. In this case, the high LDOS states in the path of the entry and exit of electrons in the system suppresses the participation of electrons in their related properties. Figure 3 C shows the effect of an electric field of $E_x = 0.15$ V/nm on the LDOS distribution. This field cause to break the symmetric LDOS distribution. Note that for this electric field, if one assumes the puckered structure of the phosphorene as two adjacent layers, the LDOS is not equally distributed in these layers.

The path for moving an electron between the two leads is not straight, since high LDOS atoms are surrounded by relatively low LDOS ones, making it an uneven path. It should be noted that atoms at the lowes x coordination, do not experience no electric potential, but the ones at the highest +x coordination, experience the highest electric potential, so the atoms between these hypothetical lines feel a relative electric potential respect to their position.

For $E_x = 0.3$ V/nm, the LDOS distribution changes, according to Figure 3 D . This field suppress the transmission probability by localizing the states more than the weaker $E_x$ field, as evidence in Figure 2 A , by the green area. High LDOS values are concentrated on the +x and -x coordination of the PDisk, meaning the electronic contribution is not favorable in these regions in the transmission probability.

The LDOS in the presence of perpendicular electric fields of 0.15 and 0.3 V/nm are presented, respectively, in panels E and F of Figure 3. For the electric fields, the LDOS distribution is mirror symmetric respect to the center, similar to the system without applying an electric field.

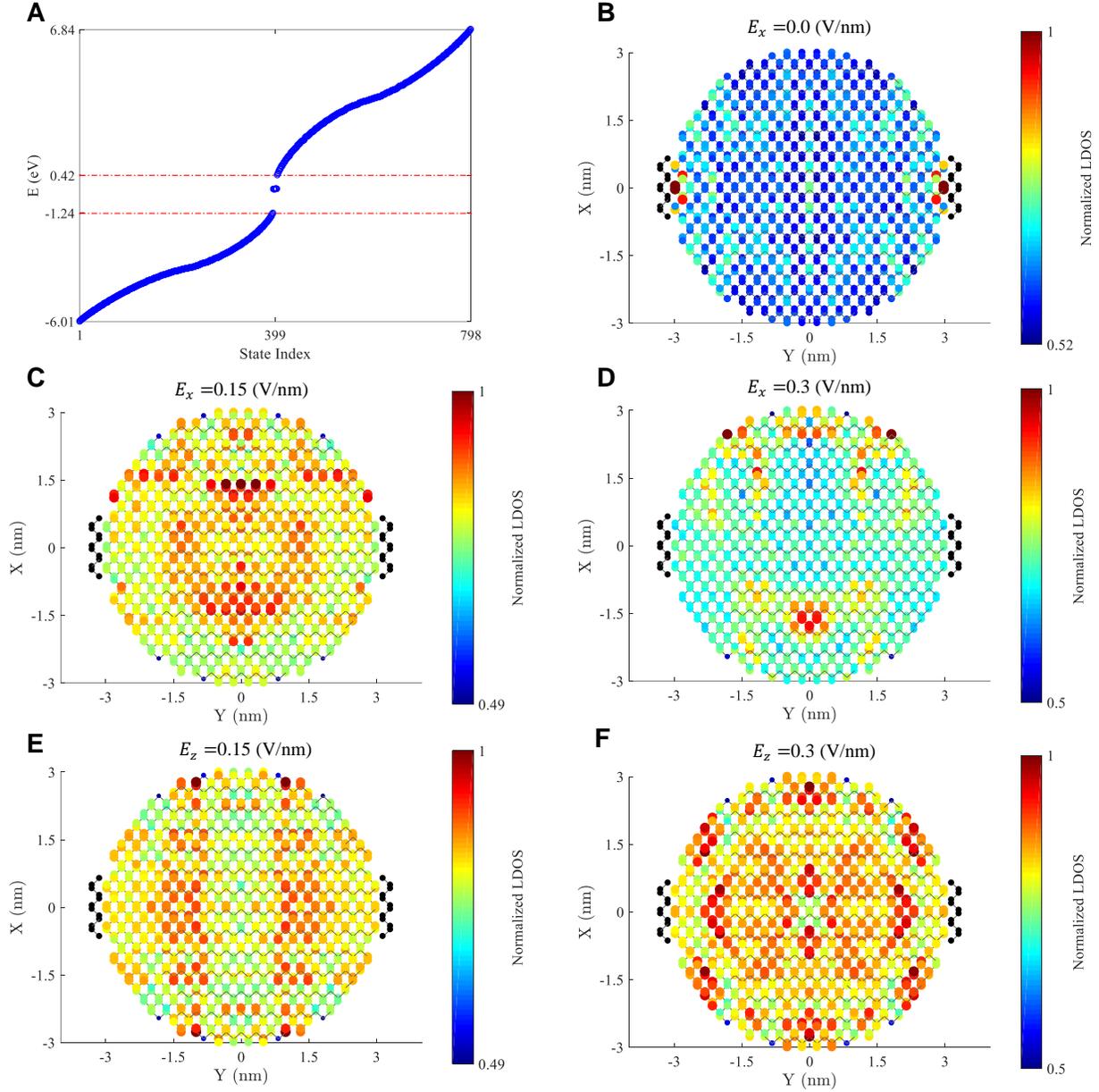

Figure 3. (A) Energy levels vs. state index for the PDisk. (B) Normalized LDOS for the PDisk in the absence of an electric field, (C) $E_x = 0.15$ V/nm, (D) $E_x = 0.3$ V/nm, (E) $E_z = 0.15$ V/nm, (F) $E_z = 0.15$ V/nm. Black dots indicate the closest unit cell atoms of the leads to the device. The gray lines are schematic of the bonding between phosphorene atoms. The bonds between the leads and the device are not drawn.

The Seebeck coefficient is plotted in Figure 4 for the leads and the transport system and in the presence of the different applied electric fields. As we mentioned earlier, more suppression of the overall transmission coefficient, especially close to the Fermi level of the system can lead to a higher Seebeck coefficient.

The Seebeck coefficient of the leads is small, which is a consequence of metallicity (thick red line in both A and B panels in Figure 4). For an applied transverse electric field (blue line in Figure 4 A ), one can see a moderate electric field, $E_x$=0.15 V/nm, suppresses S in comparison to the unbiased system, but a strong field, $E_x$=0.3 V/nm ,makes the absolute value of S slightly higher than the unbiased one.

This behavior is also evident in the case of a perpendicular electric field (Figure 4 B ), although S doesn't reach to one of no applied electric field. The sign change of the Seebeck coefficient is attributed to the change of the carrier type, meaning the high S for $E_x$=0.3 V/nm is caused by holes. These results suggest that the effect of a perpendicular electric field on S is weaker than the transverse field.

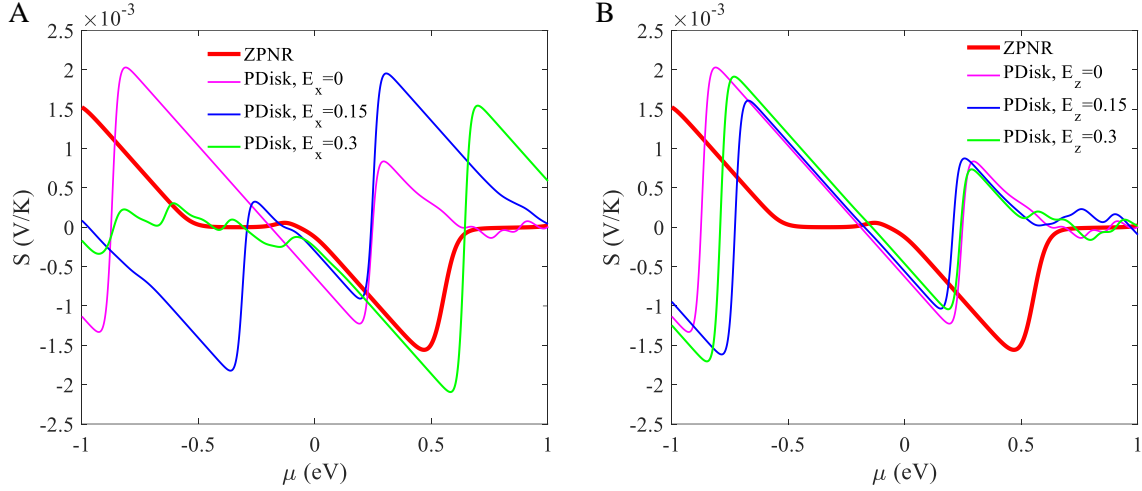

Figure 4. The Seebeck coefficient behavior vs. chemical potential for the 12-ZPNR and various electric field strengths applied along (A) the x-axis and (B) the z-axis of the PDisk. Coordinate axes are shown in Figure 1. The temperature is 300 K. Unit for the electric fields is V/nm.

From a thermoelectric point of view, the square of the Seebeck coefficient is important when calculating the thermoelectric efficiency [35]. The maximum absolute value of S in the absence of an electric field is ~2.03 mV/K at a chemical potential ($\mu$) of −0.8 eV. The highest S, 2.09 mV/K, is achieved for $E_x$ =0.3 V/nm at $\mu$ =0.58 eV. The effect of $E_z$ on S is weaker, and doesn't reach the no applied electric field. The results are listed in Table 1.

Table 1. The Seebeck coefficient and corresponding chemical potential in the presence of different electric fields.

| $E_x$ (V/nm) | $E_z$ (V/nm) | Maximum |S| (mV/K) | Corresponding $\mu$ (eV) |
|---|---|---|---|
| 0 | 0 | 2.03 | -0.81 |
| 0.15 | 0 | 1.95 | 0.30 |
| 0.3 | 0 | 2.09 | 0.58 |
| 0 | 0.15 | 1.61 | -0.78 |
| 0 | 0.3 | 1.91 | -0.73 |

## 4. Conclusion

In this work, we have studied the transmission coefficient and the Seebeck coefficient of a phosphorene nanodisk with a radius of 3.1 nm within the framework of the five-hopping TB model and with the help of NEGF formalism. According to our numerical results, engineering phosphorene to a nanodisk at the nanoscale increases the Seebeck coefficient to 2.03 mV/K, and induces an energy gap of 3.88 eV. Our numerical results have shown that with the increase of the magnitude of electric field, the transmission coefficient decreases, and the Seebeck coefficient changes. Moreover, an electric field perpendicular to the disk surface has a smaller impact on the Seebeck coefficient than a transverse electric field. A transverse electric field with the strength of 0.3 V/nm, enhances the Seebeck coefficient to 2.09 mV/K. These results can help in the design of phosphorene-based materials in thermoelectrics.

## Data availability

All data generated for this study are included in the manuscript.

## Author contributions

M.A. B carried out the simulations. M.A. B and F. K analyzed the data and prepared the manuscript. F. K. supervised the project and revised the final manuscript. All authors read and approved the final manuscript.

## Competing interests

The authors declare no competing interests.

Email: *khoeini@znu.ac.ir*